%% file: main.tex
\documentclass[letterpaper, 10 pt, conference]{ieeeconf}

\IEEEoverridecommandlockouts
\usepackage{color}
\PassOptionsToPackage{hyphens}{url}\usepackage{hyperref}
\usepackage{graphicx}
\usepackage{graphics} 
\usepackage{amsmath} 
\usepackage{amssymb}  
\usepackage{amsthm}
\usepackage{algorithm}
\usepackage{algorithmic}
\usepackage{mathrsfs}
\usepackage{subcaption}
\usepackage{tikz}
\usetikzlibrary {arrows.meta}

\usepackage{balance}

\newtheorem{definition}{Definition}
\newtheorem{problem}{Main Problem}
\newtheorem{subproblem}{Sub-Problem}

\allowdisplaybreaks[4]

\title{\LARGE \bf Automated Formation Control Synthesis from \\ Temporal Logic Specifications}

\author{Shuhao Qi$^{\dagger}$, Zengjie Zhang$^{\dagger}$, Sofie Haesaert, Zhiyong Sun
\thanks{$\dagger$ The authors contributed to this paper equally.}
\thanks{This work was supported by the European project SymAware under the grant No. 101070802, and by the European project COVER under the grant No. 101086228.}
\thanks{S. Qi, Z. Zhang, S. Haesaert, Z. Sun are with the Department of Electrical Engineering, Eindhoven University of Technology, The Netherlands.
        {\tt\small \{s.qi, z.zhang3, s.haesaert, z.sun\}@tue.nl}}
}

\input{commands}
\begin{document}

\maketitle
\thispagestyle{empty}
\pagestyle{empty}

\begin{abstract}
In many practical scenarios, multi-robot systems are envisioned to support humans in executing complicated tasks within structured environments, such as search-and-rescue tasks. We propose a framework for a multi-robot swarm to fulfill complex tasks represented by temporal logic specifications. Given temporal logic specifications on the swarm formation and navigation, we develop a controller with runtime safety and convergence guarantees that drive the swarm to formally satisfy the specification. In addition, the synthesized controller will autonomously switch formations as necessary and react to uncontrollable events from the environment. The efficacy of the proposed framework is validated with a simulation study on the navigation of multiple quadrotor robots.
\end{abstract}

\section{INTRODUCTION}
\label{sec:intro}

Multi-robot system control with complex tasks is challenging due to the high dimensions and the substantial amount of constraints~\cite{chen2022}. Taking the search-and-rescue task as an example, the robots have to sequentially achieve a series of subtasks, including locating the survivors, navigating to the rescue spots, and transporting the survivors to a designated safe zone. In this type of complex task, robot swarms should autonomously change formations to cooperatively execute sensing and communication operations or to traverse narrow spaces. Such practical tasks involve complex navigational requirements imposed by the sequential subtasks in combination with formation control requirements on the robots. While the former navigational requirements have been successfully solved by first specifying them with Linear-time Temporal Logic (LTL) and using related tools~\cite{calin2013ijrr, 2009tro}, there is not much work that explicitly incorporates the formation requirements. This paper studies the control synthesis of multi-robot systems for complex tasks represented by LTL specifications with automated formations.

For constrained control of multi-robot systems, problems such as obstacle avoidance and reachability in complex environments can be solved with recently developed control methods that also give runtime guarantees. These methods include control barrier functions (CBF) \cite{ames2017} and finite-time control Lyapunov function (CLF) ~\cite{FxT2022}. CBFs can ensure the strict satisfaction of state-dependent constraints for dynamic systems by imposing the set invariance property~\cite{ames2017}. In addition, fixed-time CLF~\cite{FxT2022}  can ensure that the system converges to a given equilibrium point or a set within user-defined time. Also, in practice, it is convenient to incorporate CBF and CLF constraints in optimization-based controllers, like quadratic programming (QP), which can balance safety and convergence without massive computation~\cite{Ames2017tac, FxT2022}. 
In this sense, CBF-CLF QP has been employed to ensure the safety and convergence of multi-robot systems~\cite{wang2017safety, tan2021distributed} and the satisfaction of the spatio-temporal constraints of temporal logic specifications~\cite{lindemann2018control, Sri2021}.

LTL specifications have been used to enable the collective behavior of multi-robot systems~\cite{calin2013ijrr, sun2022multi,calin2007tro}.  The allocation numbers of robots or systems to specific regions in space and time can be represented by counting temporal logic~\cite{tro2020CTL} and graph temporal logic~\cite{djeumou2020rss}. Alternatively, when each agent in a system has been individually assigned a distinct temporal logic specification, local cooperation of agents has been developed in \cite{lindemann2019control} where the least violating control is used in case of conflicting specifications. In contrast, in \cite{calin2007tro}, the collective behavior of multi-robot systems is achieved, where the mean and variance features of the robot swarm are employed to control arbitrarily large swarm systems to satisfy temporal logic specifications. In some practical problems, multi-robot systems are expected to not only exhibit simple collective behaviors but also to achieve specific formations to traverse special terrain and to execute sensing and communication tasks. In contrast to \cite{calin2007tro}, we aim at the explicit specification for automated formations in the navigation control of a robot swarm.

More precisely in this paper, we propose a formal framework for a multi-robot swarm system to solve a complicated autonomous navigation task with automated formations. A linear temporal logic (LTL) formula is used to specify this complicated navigation task and a symbolic model is abstracted to characterize the behavior of the swarm system. Based on this model, a symbolic controller is synthesized to generate the waypoints and desired formations subject to the LTL specification incorporating the influence of the environmental signals. Then, a QP-based control refinement method with CBFs and fixed-time CLFs is developed to ensure the satisfaction of the LTL specification with runtime guarantees. In such a way, the developed control method enables the swarm system to succeed in the navigation task with automated formations. Such a framework holds vast potential for real-world robotic swarm applications. The main contributions of this paper include 1) a framework for autonomous navigation and formation switching of multi-robot systems under LTL specifications, including finite abstractions, symbolic control synthesis, and control refinement; 2) an efficient QP-based control refinement with runtime guarantees on the task specification (autonomous navigation with automated formations) and collision avoidance.


The rest of the paper is organized as follows. Sec.~\ref{sec:fps} gives the preliminary knowledge and 
the main problem. 
The main results are given in Sec.~\ref{sec:frame}.
In Sec.~\ref{sec:cas}, we validate our framework and solutions with a simulation case on robot swarm navigation. Finally, Sec.~\ref{sec:con} concludes the paper.

\section{Preliminaries and Problem Statement}
\label{sec:fps}
%

\subsection{Multi-Robot System}
\label{sec:mrs}
Consider a multi-robot swarm system with $r $ 
robots, where each robot is described by  the following dynamic equation,
\begin{equation}
\mathcal{R}_i: \dot{x}_i(t) = g(x_i(t), u_i(t)), \quad i=1,\ldots,r,
\label{eq:sys_pb}
\end{equation}
where $x_i(t)\!\in\!\mathbb{X}\!\subset\!\mathbb{R}^n, u_i(t)\!\in\!\mathbb{U}\!\subset\!\mathbb{R}^m$ are the state and the control input of the $i$-th robot at time $t\!\in\!\mathbb{R}_{\geq 0}$, respectively, and $g\!:\!\mathbb{R}^n\!\times\!\mathbb{R}^m\!\rightarrow\!\mathbb{R}^n$ is a smooth function that describes the dynamic model of the robot. For brevity, we use vectors $x(t)=[\,x_1^T(t), x_2^T(t), \cdots, x_r^T(t)\,]^T\!\in\!\mathbb{X}^r\!\subset\!\mathbb{R}^{rn} $ and $u(t)=[\,u_1^T(t), u_2^T(t), \cdots, u_r^T(t)\,]^T\!\in\!\mathbb{U}^r\!\subset\!\mathbb{R}^{rm} $ to denote the state and control input of all robots in the system. Especially, the initial system state is denoted by $x(0) = [\,x_1^T(0), x_2^T(0), \cdots, x_r^T(0)\,]^T$.

In this paper, the interaction among the robots is described by an undirected and completed graph, in the sense that all robots in the swarm are fully connected by local communication.
We define $x_c = \frac{1}{r} \sum_{i=1}^r x_i$ as the \textit{centroid}, or the geometric center of the swarm and $x_{ij} = x_{i}-x_{j}$ as the relative displacement between two robots $i \neq j$. A certain \textit{formation} of the swarm system is described by all the relative displacements, i.e., $f = \{f_{ij}\}_{\frac{r(r-1)}{2}}$, where $f_{ij} \in \mathbb{R}^n$, for $i,j\in\{1,2,\cdots,r\}$, $i\neq j$, denotes the desired displacement between two robots. We define $\mathbb{F}$ as the formation space that contains all possible formations $f$ of the swarm system. Similarly, we define $\mathbb W \subset \mathbb{R}^n$ as the workspace of the swarm that contains all possible positions $w$ of the centroid $x_c$.

\subsection{LTL Specification for Navigation Task}
\label{sec:ltl}

We first introduce Linear Temporal Logic (LTL)~\cite{modelchecking2008, calin2017book} as follows. \\
\noindent{\textbf{Syntax.}} The syntax of LTL is recursively defined as,
\begin{equation}
    \psi ::= \top \mid p \mid \neg \psi \mid \psi_1 \wedge \psi_2 \mid \bigcirc \psi \mid \psi_1 \mathsf{U} \psi_2,
\end{equation}
where  $\psi_1$, $\psi_2$ and $\psi$ are LTL formulas, $p \in \AP$ is an atomic proposition, $\neg$ is the negation operator, $\wedge$ is the conjunction operator that connects two LTL formulas, and $\bigcirc$ and $\mathsf{U}$ represent the \textit{next} and \textit{until} temporal operators, respectively. Based on these essential operators, other logical and temporal operators, namely \textit{disjunction} $\vee$, \textit{implication} $\rightarrow$, \textit{eventually} $\lozenge$, and \textit{always} $\square$ can be defined as, $\psi_1 \vee \psi_2:= \lnot \!\left(\psi_1 \wedge \psi_2 \right)$, $\psi_1 \rightarrow \psi_2$ $:= \lnot \psi_1 \vee \psi_2$, $\lozenge \psi :=\top \mathsf{U} \psi$, and $\square \psi :=\neg \lozenge \neg \psi$.
\smallskip

\noindent{\textbf{Semantics.}} Consider a set of atomic propositions $\AP = \left\{ p_1, \dots, p_N \right\}$ which defines an alphabet $2^{\AP}$, where each letter $\omega \in 2^{\AP}$ contains the set of atomic propositions that are true. An infinite string of letters is a word $\pmb{\omega} = \omega_0 \omega_{1} \omega_{2} \dots$, where $\omega_i \in 2^{\AP}$, $i \in \mathbb{N}_{\geq 0}$, with a suffix $\pmb{\omega}_k = \omega_k \omega_{k+1} \omega_{k+2} \dots$, $k \in \mathbb{N}_{\geq 0}$. For a given word $\pmb{\omega}$, basic semantics of LTL are given as
$\pmb\omega_k \models p$, if $p \in \omega_k$; $\pmb\omega_k \models \lnot p$, if $p \notin \omega_k$;
$\pmb\omega_k \models \psi_1 \wedge \psi_2$, if $\pmb \omega_k \models \psi_1$ and $\pmb \omega_k \models \psi_2$;
$\pmb\omega_k \models \bigcirc \psi$, if $\pmb\omega_{k+1} \models \psi$;
$\pmb\omega_k \models \psi_1 \mathsf{U} \psi_2$, if $\exists$ $i \in \mathbb{N}$ such that $\pmb\omega_{k+i} \models \psi_2$, and $\pmb\omega_{k+j} \models \psi_1$ holds $\forall \, 0\leq j < i$.
\smallskip

\noindent{\textbf{Reactive LTL formulas for robot swarm.}}
In this paper, we use an LTL formula to specify whether the swarm system achieves the navigation task with automated formation. Motivated by~\cite{calin2007tro}, we select to use the centroid position and formation as essential features to respectively characterize the navigation and interaction behaviors of the swarm system. Therefore, we introduce atomic propositions for the swarm centroid in the workspace $\AP_w$ based on a labeling function $\mathcal{L}_w\!:\!\mathbb W\!\rightarrow\!2^{\AP_w}$ and those for the formation $\AP_f$ with a labeling function $\mathcal{L}_f\!:\!\mathbb F \!\rightarrow\!2^{\AP_f}$. Moreover, to specify how the swarm system should react to external signals from the environment, we use a finite set $\mathcal{E}$ to denote all possible states of the environmental signal and the set of atomic propositions $\AP_e$ to describe the property of the signal. Their correspondence is described by a labeling mapping $\mathcal{L}_e\!:\!\mathcal{E}\!\rightarrow\!2^{\AP_e}$. In this sense, the overall set of atomic propositions of the swarm system for the navigation task is $\AP\!:=\!\AP_e\cup\AP_w\cup\AP_f$. Then, we can use an LTL formula $\psi$ defined on the output word of the swarm system $\pmb{\omega} = \omega_0 \omega_{1} \omega_{2} \dots$, where $\omega_k\!\in\!2^{\mathsf{AP}}$, $k\!\in\!\mathbb{N}$, to specify the navigation task. We say that the robot achieves the navigation task if its output word satisfies the specification, i.e., $\pmb{\omega}\vDash \psi$. In this sense, the overall labeling mapping of the swarm system is the combination of all mappings $\mathcal{L}_w$, $\mathcal{L}_f$, and $\mathcal{L}_e$, i.e., $\mathcal{L}: \mathbb{W} \times \mathbb{F} \times \mathcal{E} \rightarrow 2^{\AP}$. The subset LTL formulas known as Generalized Reactive(1) (GR(1))~\cite{BLOEM2012911}, is specifically well fitted to deal with both external and internal variables. GR(1) formulas are of the form  $\psi:= \psi_e \rightarrow \psi_s$, where $\psi_e$ constrains the allowed behavior of the uncontrolled propositions  $\mathsf{AP}_e$ and $\psi_s$ constrains the desired behaviors of the system with the controlled propositions $\mathsf{AP}_w \cup\mathsf{AP}_f$.


\subsection{Problem Statement}\label{sec:probstate}

Given the multi-robot system in \eqref{sec:mrs} and the task specification $\psi$ defined in \eqref{sec:ltl}, the navigation task studied in this paper can be formulated as a synthesis problem for the specification $\psi$. Before we give the formal problem statement, we explain how automated formation is addressed for the navigation task. The main objective of the navigation task is that the centroid of the swarm system should ultimately reach the navigation goal along a feasible path in the environment (\textit{navigation}). Meanwhile, the environmental terrain requires that certain spots must be passed by certain formations. Therefore, the swarm system should also automatically switch to the feasible formations when passing the corresponding spots (\textit{automated formation}). Also, the swarm system should always react to the environmental signals and perform the correct response (\textit{reaction}). Besides, all robots in the swarm should avoid collisions with each other and the obstacles in the environment (\textit{collision avoidance}). The formal problem statement is given as follows.



\begin{problem}[Navigation with Automated Formation]
\label{pb:overall}
For a multi-robot swarm system $\mathcal{R} = \{\mathcal{R}_1, \mathcal{R}_2, \cdots, \mathcal{R}_r\}$ as defined in Eq.~\eqref{eq:sys_pb} with a state space $\mathbb{X}^r$, a workspace $\mathbb{W}$, a formation space $\mathbb{F}$, an environmental signal space $\mathcal{E}$, a labeling mapping $\mathcal{L}:\mathbb{W}\times\mathbb{F}\times\mathcal{E} \rightarrow \mathsf{AP}$, and a GR(1) formula $\psi:= \psi_e \rightarrow \psi_s$ defined on the alphabet $2^{\AP}$ specifying the requirements on navigation, automated formation, reaction, and collision avoidance, find a provably-correct reactive control policy to ensure that the output word $\pmb{\omega}$ satisfies the specification $\psi$, i.e, $\pmb{\omega} \vDash \psi$. \qed
\end{problem}
\section{Framework and Controller Design}
\label{sec:frame}

\subsection{The Control Design Framework }
To capture behaviors of the dynamic system in Eq.~\ref{eq:sys_pb}, an abstract system can be defined in the following formulation,
\begin{definition}\label{def:fts}
(Deterministic Finite Transition System, DFTS):
A deterministic finite transition system is a tuple $\mathcal{T}\!=\! (S, s_0,  A, \delta$, $\AP$, $\mathcal{L})$, where $S$ and $A$ are finite sets of states and actions, $s_0$ is the initial state, $\delta\!:\! S \!\times\! A \!\rightarrow\! S$ is a transition function that prescribes the state transition under a certain input, $\AP$ is a finite set of atomic propositions, and $\mathcal{L}\!:\! S \!\rightarrow\! 2^{\AP}$ is a labeling function. \qed
\end{definition}

	Given a sequence of actions $\mathbf{a} \!=\! \ac_0 \ac_1 \ac_2 \cdots$ with $\ac_i \!\in\! \A$, 
	a DFTS $\mathcal{T}$ initiated at $s_0 \!\in\! S$ generates a trajectory or a run $\pmb{s} \!=\! s_0 s_1 s_2, \ldots$, where $s_{i+1}\!=\!\delta(s_i,\ac_i)$, $i \!\in\! \mathbb{N}$. Accordingly, the output word of the DFTS $\pmb \omega = \omega_0 \omega_1 \omega_2 \cdots$ is uniquely defined for a given initial state $s_0 \in S$, where $\omega_i \in 2^{\AP}$, $i \in \mathbb{N}$.
Let a history $h_k = s_0 a_0 s_1 a_1  s_2 a_2...s_k$ be given at time $k$ with $h_k\!\in\! H$, we can then define a control strategy as a map $\pia$ from the set of histories to the action set, i.e., $\pia: H \!\rightarrow\! \A$.
A control strategy in the form  $\pia\!:\! S \!\rightarrow\! \A$ is a Markov strategy as it only depends on current state of $\mathcal{T}$.

\smallskip

In the next subsection, we will show that we can exactly develop a symbolic model as the abstraction of the swarm system in Eq.~\eqref{eq:sys_pb} using a DFTS with the same atomic proposition sets $ \mathsf{AP}_w$ and $\mathsf{AP}_f$. 
More specifically, let us denote two finite sets $\mathcal{W}\!:=\!\{w_1,w_2,\ldots\}\!\subset\!\mathbb W$ and $\mathcal{F}\!:=\!\{f_1,f_2,\ldots\}\!\subset\!\mathbb F$ with waypoints in the workspace $\mathbb W$ and the formation space $\mathbb F$ of the swarm system. 
The state set of the DFTS $S\!:=\mathcal{W}\!\times\!\mathcal{F}$ is an abstraction of the feature space of the swarm system $\mathbb{W}\times \mathbb{F}$. As a consequence, the task specification $\psi$ defined for the swarm system can now be specified for the abstract model with labeling map ${\mathcal{L}}\!:\!\mathcal{W}\!\times\!\mathcal{F} \!\times\!\mathcal{E} \!\rightarrow\!2^{\AP}$. Consider the history signal $h_k^{\mathcal{E}}\!:=\!s_0 e_0 a_0 s_1 e_1 a_1  s_2e_1 a_2...s_k\!\in\! H^{\mathcal{E}}$ extended with the uncontrolled variables $e_k\in \mathcal E$. Then, we can decompose the main problem into two sub-problems with one being the control synthesis for the abstract model $\mathcal{T}$ and the other as control refinement to guarantee the soundness of the output words $\pmb{\omega}$ for the concrete swarm system.

\begin{subproblem}[Symbolic control synthesis]
\label{spb:strategy}
Consider a DFTS $\mathcal{T}$ defined in Def.~\ref{def:fts} as the abstraction of the swarm system given in Eq.~\eqref{eq:sys_pb}. Synthesize a symbolic control strategy, $\pia: H^\mathcal{E} \rightarrow A$, such that for any feasible environmental signal from $\mathcal{E}$ and for the initial state $s_0$, the LTL specification $\psi$ is satisfied.
\end{subproblem}


Given the control strategy $\pia$ for the abstract system $\mathcal{T}$, we should also make sure there is a refined control policy for the swarm system that drives the multi-robot system to satisfy the navigation specification $\psi$ without collisions inside and outside the swarm.




\begin{subproblem}[Control refinement]
\label{spb:control}
For the swarm system given in Eq.~\eqref{eq:sys_pb} and its abstract model, a DFTS $\mathcal{T}$ defined in Def.~\ref{def:fts}, with a symbolic control strategy $\pia$ solved in \textit{Sub-Problem~\ref{spb:strategy}}, design a controller that maps the symbolic state action pairs to a continous control map 
 $ \pi:\X^r\times S \times A\rightarrow \mathbb U^r$,
such that the output word of the swarm system $\pmb{\omega}$ assigned by the labeling mapping $\mathcal{L}$ satisfies the LTL specification defined in Sec.~\ref{sec:ltl}, i.e., $\pmb{\omega} \models \psi$.
\qed
\end{subproblem}


\subsection{Symbolic Control Synthesis}
\label{sec:scs}
This subsection discusses the synthesis of the symbolic control strategy in \textit{Sub-Problem~\ref{spb:strategy}}. The synthesis process is performed in three steps:
 
\subsubsection{Step 1: Determining Waypoint Sets}
The first step is to determine the sets of waypoints in the workspace and the formation, namely $\mathcal{W}$ and $\mathcal{F}$ which are important to construct the abstraction model $\mathcal{T}$, as introduced in Sec.~\ref{sec:frame}.
A waypoint is a desired position $w\!\in\!\mathcal{W}$ for the swarm centroid to reach or a desired formation $f\!\in\!\mathcal{F}$ for the swarm to achieve.
In this work, the waypoints are selected by fully incorporating the physical property of the swarm, the practical conditions of the environment, and the specific requirements of the tasks.  For example, there might exist some narrow spaces where the swarm can only pass with a certain formation. Also,  the number of the waypoints is selected as possibly small while maintaining a dense distribution to make a balance between the scale of the abstraction model and the feasibility of the navigation task.

\subsubsection{Step 2: Realization of the DFTS}
After determining the finite sets of waypoints $\mathcal{W}$ and $\mathcal{F}$, the next step is to use them to construct a DFTS $\mathcal{T}\!=\!\{S, s_0, A, \delta, \AP, \mathcal{L}\}$ to realize an abstract model for the swarm system. Directly, the state set is $S\!=\!\mathcal{W}\!\times\!\mathcal{F}$. The action set $A$ is also a finite symbolic set that contains all possible actions for state transitions. The atomic proposition set $\mathsf{AP}$ and labeling mapping $\mathcal{L}$ are the same as Sec.~\ref{sec:ltl}.
Thus, the most critical part of this step is to determine the transition function $\delta$ which describes to which waypoint the system transits given the current waypoint. We first assume a full transition and then eliminate the infeasible transitions by looking into the environmental conditions and the dynamic model of the swarm system.
%
%
For example, we eliminate a transition from one waypoint to another if it forces the swarm to pass an area with an impractical formation.
Moreover, the transition should also incorporate the feasibility of the control refinement process described by \textit{Sub-Problem~\ref{spb:control}}, which will be discussed in the next subsection.

\subsubsection{Step 3: Synthesis of  Control Strategy}
Given the abstract, symbolic model, we can use an off-the-shelf tool to solve the control synthesis problem. More precisely, we use the Omega solver \cite{filippidis2016symbolic}
in Tulip~\cite{wongpiromsarn2011tulip,tulip2016}. Internally, it constructs an enumerated transducer that ensures the satisfaction of the GR(1) formula for any admissible behavior of the uncontrolled variables $e_k$. For the overall synthesis procedure can be referred to~\cite{BLOEM2012911, tulip2016}. Note that if the environment variable takes a value other than the imposed assumptions, no guarantees can be provided on the system behavior.

\subsection{Control Refinement with Runtime Guarantees}
\label{sec:qp}

In this subsection, we design a QP-based control refinement to solve \textit{Sub-Problem~\ref{spb:control}}. Let the symbolic state $ s=(w,f)\in S$ and the symbolic action $a\in A$ be given for which the next symbolic state should be $s_+ = \delta(s,a)$. The refined control inputs $u(t)$ should control the robot swarm such that the desired waypoints $(w,f)$ defined by the symbolic control are reached. Additionally, different constraints are added concerning the runtime safety requirements.

\subsubsection{Computing Robot Control Inputs Via Solving QP}

Without losing generality, we consider a swarm with single integrator models, i.e., $\dot{x}_i(t) \!=\! u_i(t)$  for $i\!\in\!\{1,2,\cdots,r\}$. For \textit{Sub-Problem~\ref{spb:control}}, we need to solve the following QP problem formulated in Eq.~\eqref{eq: swarm_qp}, for all $i \!\neq\! j$, $i,j \!\in\! \{1,2,\cdots,r\}$,
\begin{subequations}
\vspace{0.1cm}
\label{eq: swarm_qp}
\begin{align}
\min _{z}  \textstyle  z^{\mathrm{T}} H z  +Q^{\mathrm{T}} z & \\
\text { s.}\text{t. } \quad \left\| u_i \right\| \leq\, u_{\max}&, \label{subeq: limit}\\
\begin{split}
\frac{\partial h_{\mathcal{W}}(x_c, w)}{\partial x_c}  u_c  \!\leq&\,   \delta_1 h_{\mathcal{W}}(x_c,w)\\
&\textstyle - \alpha_1\! \max^{\gamma_1} \{0, h_{\mathcal{W}}(x_c, w)\}\\
&\textstyle -\alpha_2\! \max^{\gamma_2} \{0, h_{\mathcal{W}}(x_c, w)\},
 \label{subeq: center}
\end{split} \\
\begin{split}
\frac{\partial h_{\mathcal{F}}(x_{ij}, f_{ij})}{\partial x_{ij}}  u_{ij} \!\leq&\, \delta_1 h_{\mathcal{F}}(x_{ij}, f_{ij}) \\
&\textstyle -\alpha_1 \!\max^{\gamma_1} \{0, h_{\mathcal{F}}(x_{ij}, f_{ij})\} \\
&\textstyle -\alpha_2 \!\max^{\gamma_2} \{0, h_{\mathcal{F}}(x_{ij}, f_{ij})\},
\label{subeq: form}
\end{split} \\
 \frac{\partial h_{\mathcal{D}}(x_{ij})}{\partial x_{ij}} u_{ij} \geq & -\delta_2 h_{\mathcal{D}}(x_{ij}), \label{subeq: collision}\\
\frac{\partial h_{\mathcal{O}}(x_i)}{\partial x_i} u_i \geq & -\delta_2 h_{\mathcal{O}}(x_i),\label{subeq: obs}
\end{align}
\end{subequations}
where $z = [u^T, \delta^T]^T\!\in\!\mathbb{R}^{2 \times r + 2}$ are decision variables, in which $\delta\!=\![\,\delta_1,\,\delta_2\,] \in \mathbb{R}^2$ are slack variables, $H$ is a diagonal matrix with positive constant elements, $Q\!=\![\,\mathbf{0}_{2\times r}, w_{\delta_1}, 0\,]$ where $w_{\delta_1}\!\in\!\mathbb{R}^+$ is a penalizing scalar of the slack variable $\delta_1$, $u_{\max}\!\in\!\mathbb{R}^+$ defines the control limit of the system, and $u_{ij}\!=\!u_{i}-u_{j}$ for any $i,j\in\{1,2,\cdots,r\}$, denote the control difference between two robots $i\neq j$. Sublevel sets of $h_{\mathcal{W}}(x, w) \!=\! \|x-w \|^2  \!-\! d_G^2$ and
$h_{\mathcal{F}}(x_{ij}, f_{ij})\!=\!\|x\!-\!f_{ij} \|^2\!-\!d_F^2$ represent desired centroid position $w\!\in\!\mathbb W$ and swarm formation $f\!\in\!\mathbb F$ for the robot swarm respectively, where $d_G, d_F \in \mathbb{R}^+$ are tolerance thresholds. Superlevel sets of $h_{\mathcal{D}}(x)\!=\!\|x \|^2 - d_O^2$ and $h_{\mathcal{O}}(x)\!=\!1\!-\!(x\!-\!\eta)^T \!P(x\!-\!\eta)$ denote collision-free sets with other robots and local obstacle-free sets respectively, where $d_O\!\in\!\mathbb{R}^+$ is the minimal distance among agents and $\eta\!\in\!\mathbb{R}^n$ and $P\!\in \mathbb{R}^{n\!\times\!n}$ are constant parameters determined by the environment. Constant parameters $\alpha_1$, $\alpha_2$, $\gamma_1$, $\gamma_1$ are chosen as $\alpha_1\!=\!\alpha_2\!=\!\frac{\mu \pi}{2 T_{ud}}$, $\gamma_1\!=\!1\!+\!\frac{1}{\mu}$, $\gamma_2\!=\!1\!-\!\frac{1}{\mu}$ with user-defined constants $\mu>1$ and $T_{ud}$.





To achieve the desired waypoints $w$ and $f$, fixed-time CLF ensures the convergence to a given set within user-defined time $T_{ud}$, and the concrete formulation can be found in~\cite{FxT2022}. Constraints (\ref{subeq: center}) and (\ref{subeq: form}) are formulated in the form of fixed-time CLF, which respectively drive the swarm system to reach a waypoint $w \!\in\! \mathbb{W}$ and achieve a formation $f\!=\!\{f_{ij}\}_{\frac{r(r-1)}{2}}$ that corresponds to the chosen next  abstract state $(w, f)$, with $w\!\in\!\mathcal{W}$, $f\!\in\!\mathcal{F}$. As for runtime safety, CBF is an effective tool to ensure the set invariance of a dynamic system in a given set, which is often used to ensure system safety of collision avoidance. In the form of control inputs condition for CBF~\cite{ames2017}, constraints (\ref{subeq: obs}) and (\ref{subeq: collision}) attempt to keep the robots within the safety sets $\mathcal{D}$ and $\mathcal{O}$ to avoid collisions with each other and with the obstacles in the environment. This corresponds to the \textit{collision avoidance} objective in Sec.~\ref{sec:probstate}. Note that the \textit{reactivity} objective has been achieved already in the symbolic control in Sec.~\ref{sec:scs}.

\subsubsection{Specification Satisfaction Analysis}
If the QP in Eq.~\eqref{eq: swarm_qp} has a feasible solution $u(t)$ and the slack variable $\delta_1$ remains negative, the obtained control inputs $u(t)$ guarantee that the swarm reaches the given waypoint $(w,f)$ within a maximum time $T_{ud}$~\cite{FxT2022}. This indicates that, for any waypoints and desired formations that satisfy the specification $\psi$, the swarm can track them within a finite predefined timing bound. Although the swarm may traverse other positions and formations between two waypoints, it ultimately reaches the next waypoint $(w,f)$ within a strict timing.
This means that, with the tolerance of the transient states of the swarm with a strict timing bound, the behavior of the swarm satisfies the given task specification $\psi$. Moreover, the constraints~\eqref{subeq: collision} and~\eqref{subeq: obs} guarantee that the system never traverses to dangerous regions. Therefore, we can claim that \textit{Sub-Problem~\ref{spb:control}} is solved if the QP in Eq.~\eqref{eq: swarm_qp} is feasible.

\begin{figure*}[htbp]
     \centering
     \vspace{0.2cm}
     \begin{subfigure}[b]{0.18\textwidth}
         \centering
         \includegraphics[width=\textwidth]{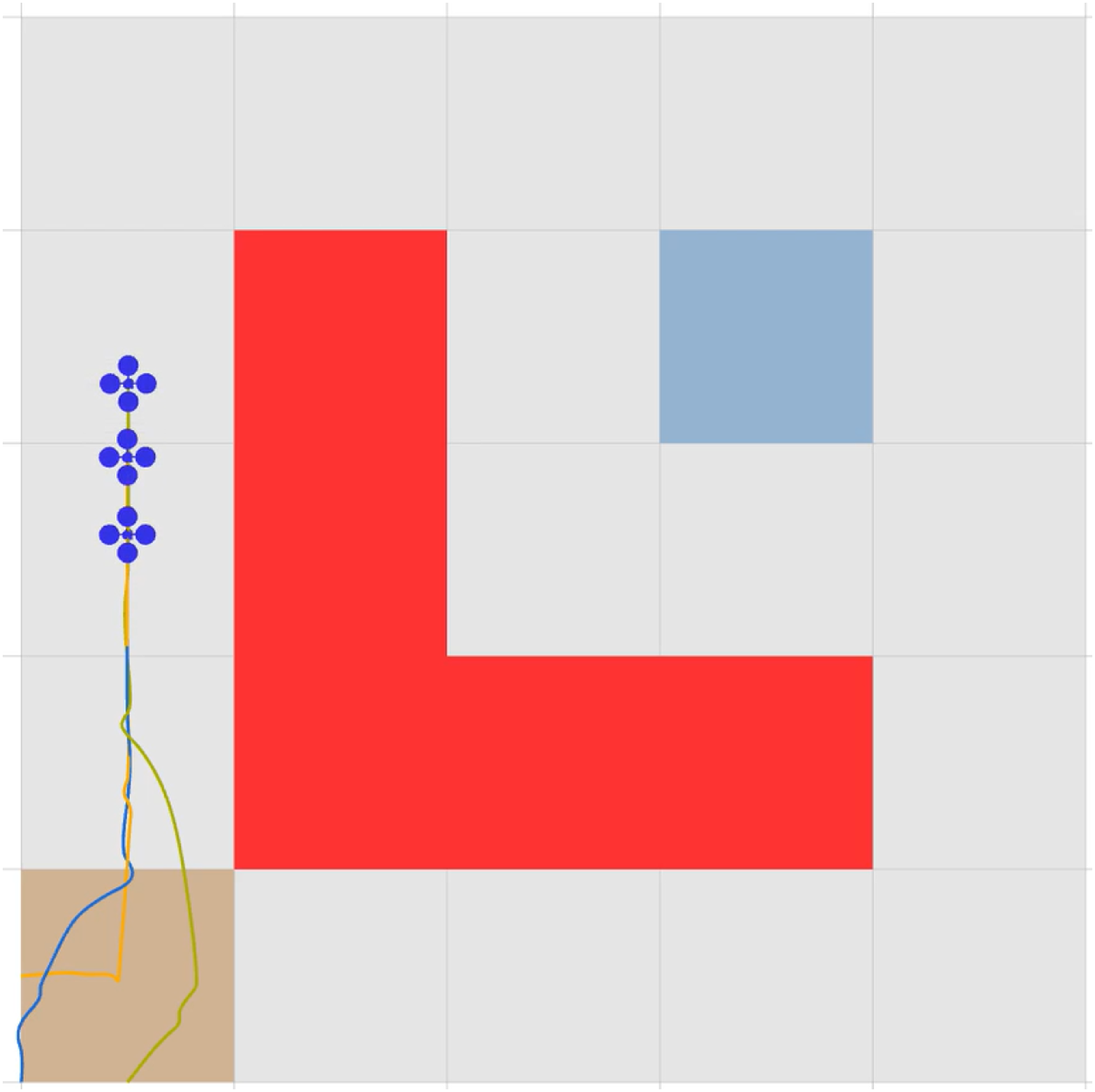}
         \caption{``battery'': True}
         \label{fig:f11}
     \end{subfigure}
     \hfill
     \begin{subfigure}[b]{0.18\textwidth}
         \centering
         \includegraphics[width=\textwidth]{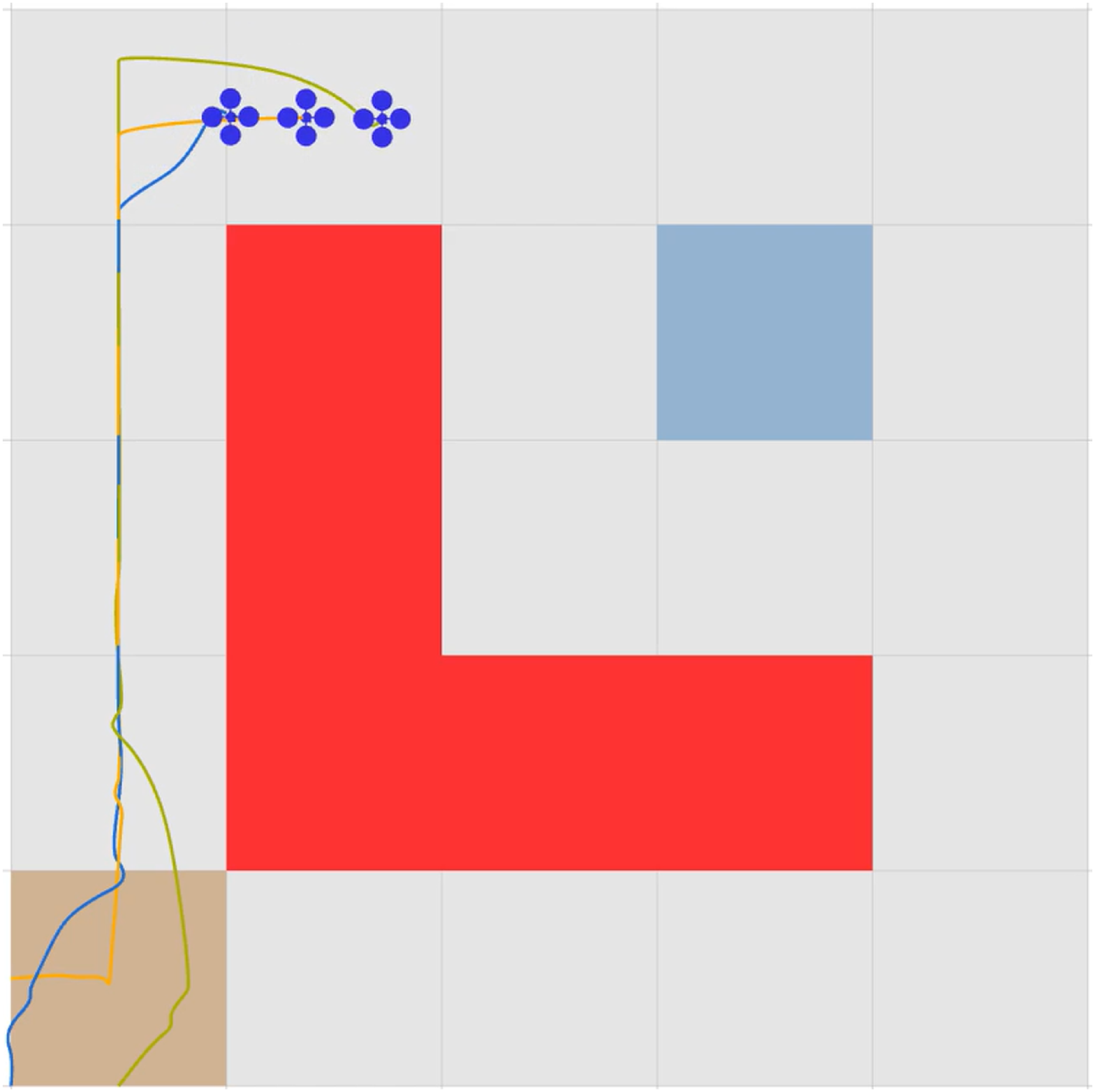}
         \caption{``battery'': True}
         \label{fig:f12}
     \end{subfigure}
     \hfill
     \begin{subfigure}[b]{0.18\textwidth}
         \centering
         \includegraphics[width=\textwidth]{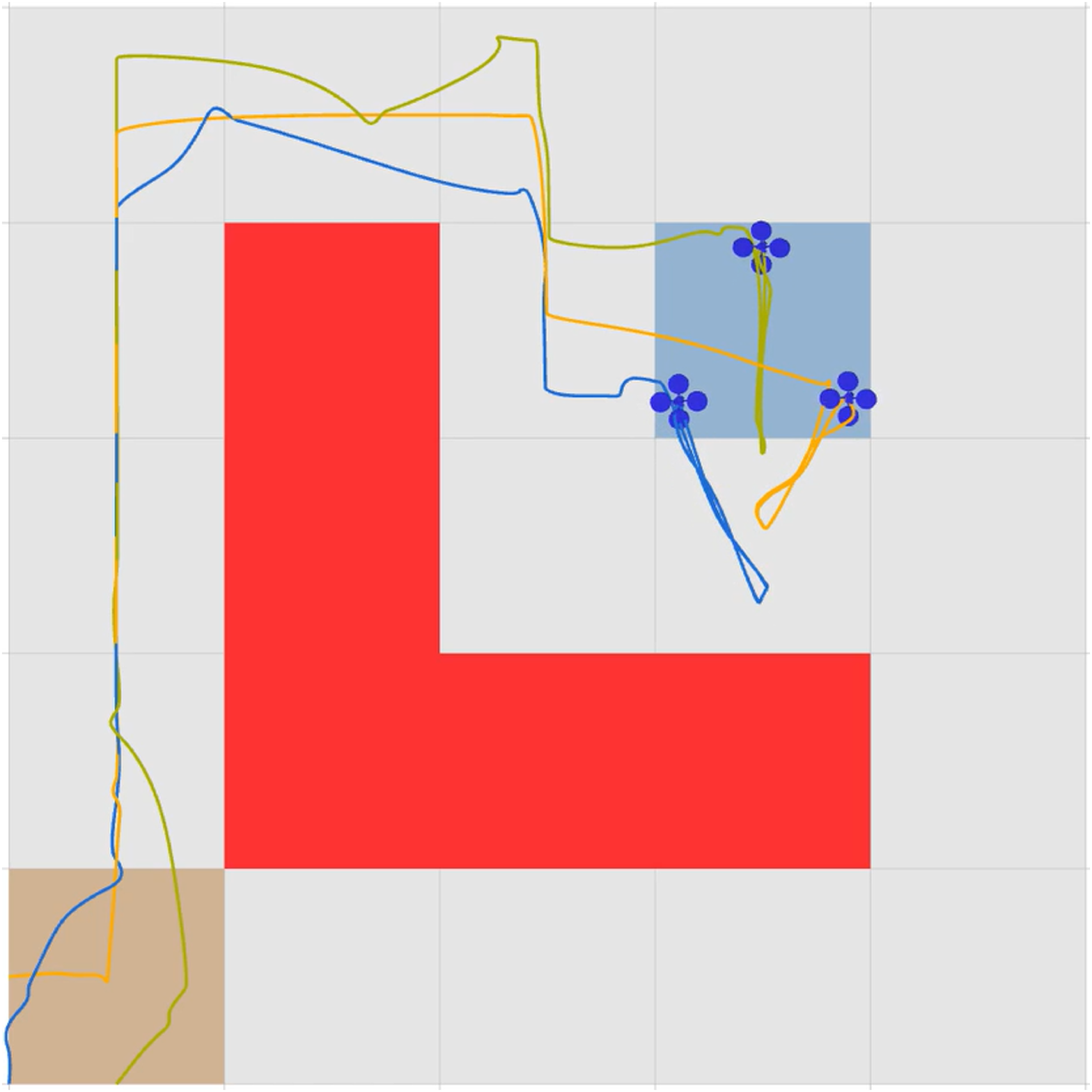}
         \caption{``battery'': True}
         \label{fig:f13}
     \end{subfigure}
     \hfill
     \begin{subfigure}[b]{0.18\textwidth}
         \centering
         \includegraphics[width=\textwidth]{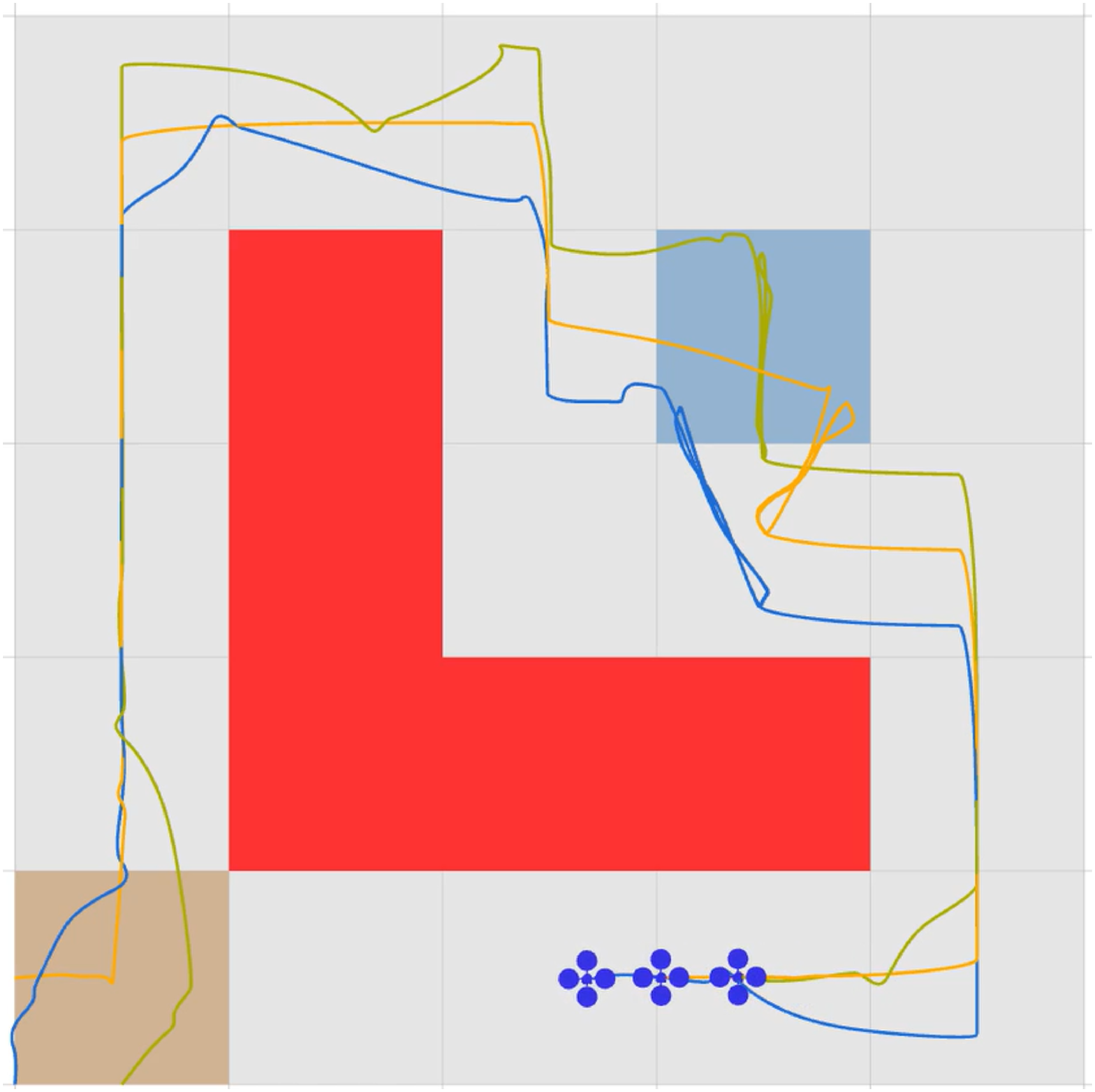}
         \caption{``battery'': False}
         \label{fig:f14}
     \end{subfigure}
     \hfill
     \begin{subfigure}[b]{0.18\textwidth}
         \centering
         \includegraphics[width=\textwidth]{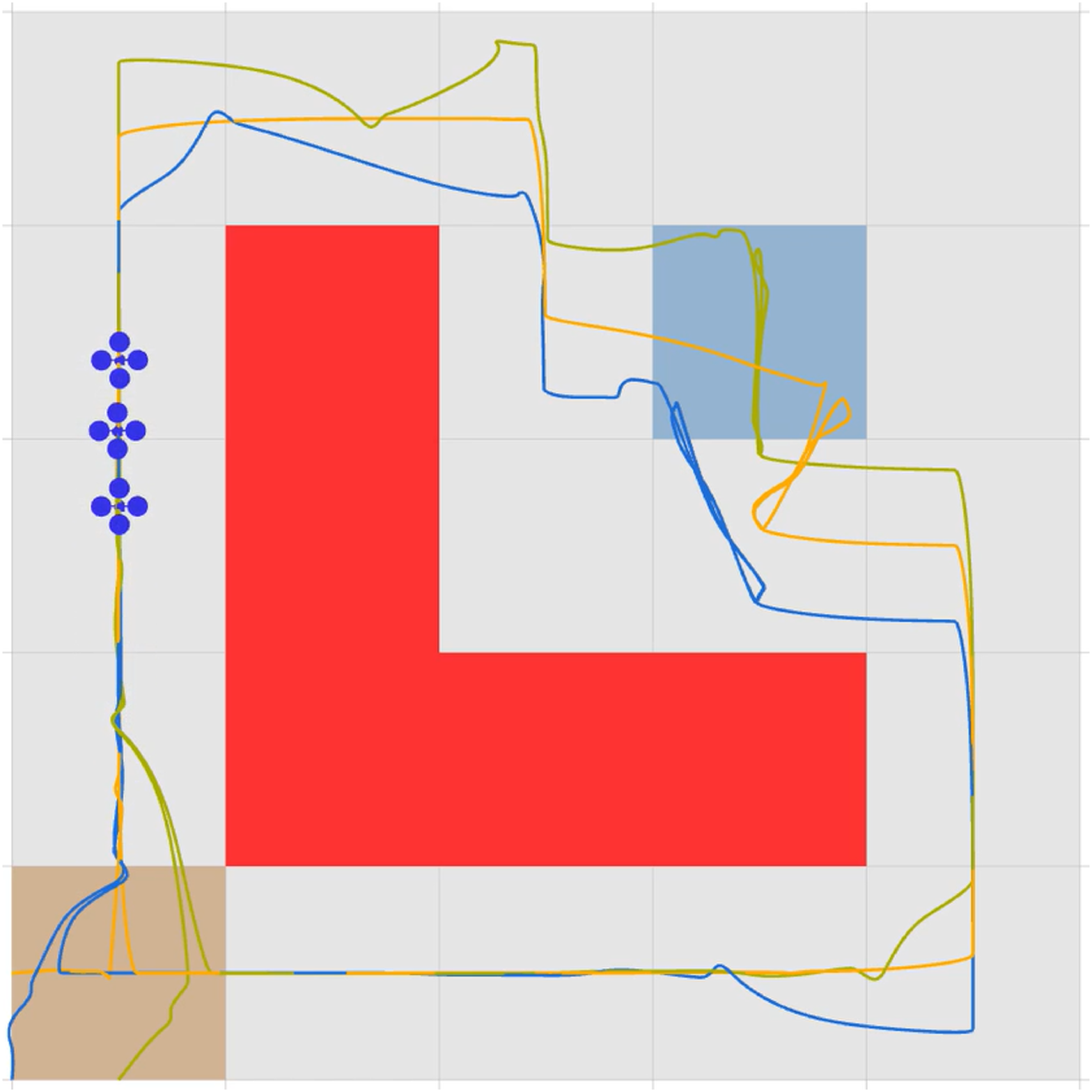}
         \caption{``battery'': True}
         \label{fig:f15}
     \end{subfigure}
        \caption{The planar view of the environment and the robot trajectories in a simulation run, as time changes (left to right).}
        \label{fig:f1}
        \vspace{-0.1cm}
\end{figure*}

If the QP in Eq.~\eqref{eq: swarm_qp} is infeasible, it might be caused by the constraints encoding the impractical waypoints or formations. In this case, the infeasibility can be resolved by pruning the transition function $\delta$ of the abstraction model $\mathcal{T}$ as developed in Sec.~\ref{sec:scs}. Specifically, the waypoints leading to the infeasible QP are recognized as infeasible transitions in the abstraction model $\mathcal{T}$ and are eliminated from $\delta$. This is iteratively performed until all transitions in a synthesized strategy ensure the feasibility of the QP.

\section{Case Study}\label{sec:cas}

In this section, we use a swarm navigation control case in simulation to validate the efficacy of our proposed framework and solution. Consider a homogeneous swarm system that contains $r=3$ quadrotor robots moving in a two-dimensional planar environment $\X \subset \mathbb{R}^2$, where $\X$ is a $5\,$m $\times5\,$m square area. The dynamic model of the $i$-th robot of the swarm system is given as the following single integrator $\dot{x}_i(t) = u_i(t)$, $i=1,2,3$, where $x_i(t), u_i(t) \in \mathbb{R}^2$ are the position and control input of robot $i$ at time $t \in \mathbb{R}_{\geq 0}$. In this setting, the workspace containing the possible positions of the centroid is also $\mathbb{X}$, i.e., $\mathbb{W}:=\mathbb{X}$.

The terrain of the environment is illustrated in Fig.~\ref{fig:f1}. The environment is split into a $5\times 5$ grid which generates 25 even square blocks, each with a size of $1\,$m $\times 1\,$m. The yellow block is the starting point of the robot swarm. The blue block is the navigation goal that the swarm needs to reach ultimately. The red blocks are the obstacles that the robots should avoid. All waypoints in $\mathcal{W}$ are assigned as the center of the accessible square partitions. Also, we determine an abstract formation set including three different formations $\mathcal{F}=\{f_1, f_2, f_3\}$ by partitioning $\mathbb F$, where $f_1$, $f_2$, $f_3$ represent a horizontal formation (as shown in Fig.~\ref{fig:f12} and Fig.~\ref{fig:f14}), a vertical formation (as shown in Fig.~\ref{fig:f11} and Fig.~\ref{fig:f15}), and a triangle shape formations (Fig.~\ref{fig:f13}), respectively. The definitions of $\mathcal{W}$ and $\mathcal{F}$ can be found in our online document on~\cite{ourCode}.

\begin{figure}[htpb]
    \centering
    \includegraphics[width=0.38\textwidth]{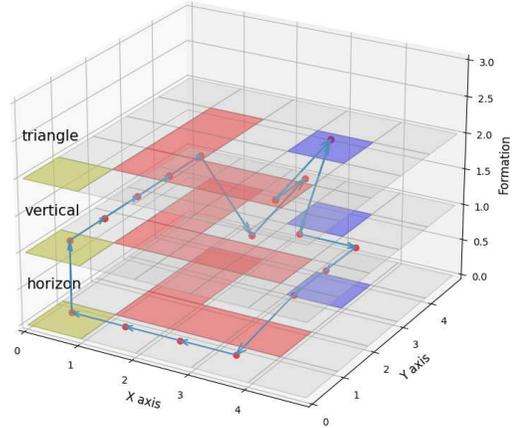}
    \caption{The visualization of the states of the abstract model. The three planes distributed along the $z$-axis are the environment with waypoints corresponding to three formations. Each small square block is an abstract state. The blue line denotes the transition of the abstract states in a simulation run. The passing waypoints are marked as red dots.}
    \label{fig:strategy}
    \vspace{-0.1cm}
\end{figure}

The DFTS $\mathcal{T}\!=\!\{S, s_0, A, \delta, \AP, \mathcal{L}\}$ as the abstract model of the quadrotor swarm is realized as follows. The state space $S \!=\! \mathcal{W} \!\times\! \mathcal{F}$, where $\mathcal{W}$, $\mathcal{F}$ are realized as finite sets with $25$ and $3$ elements, respectively. The transition relation $\delta$ is represented as a matrix and can also be found in our Github repository~\cite{ourCode}. The main principles we use to construct the transition relation are as follows.
\begin{itemize}
\item When the swarm passes by a narrow corridor, it should switch to the thinnest formation to fit its direction.
\item The swarm should not enter an obstacle (red) region.
\end{itemize}
The atomic proposition set is $\AP = \AP_w \cup \AP_f \cup \AP_e$, where $\AP_w \!=\! \{$freespace, home, goal, obstacle$\}$, $\AP_f\!=\!\{$horizon, vertical, triangle$\}$, and $\AP_e=\{$battery$\}$. The label mapping is $\mathcal{L} \!=\! \{\mathcal{L}_w, \mathcal{L}_f,\mathcal{L}_e\}$. Mapping $\mathcal{L}_w$ labels the yellow block in Fig.~\ref{fig:f1} as ``home'', the blue block as ``goal'', the red blocks as ``obstacle'', and all other blocks as ``freespace''. Mapping $\mathcal{L}_f$ is defined such that $\mathcal{L}_f(f_1) =$ ``horizon'', $\mathcal{L}_f(f_2)=$ ``vertical'', and $\mathcal{L}_f(f_3) =$ ``triangle''. Then, mapping $\mathcal{L}_e$ gives ``battery'' if all the batteries of the robots are charged. The abstract model is visualized in Fig.~\ref{fig:strategy}, where the $x$-$y$ planes along the formation axis show the planar view of the environment for different formations. Thus, Fig.~\ref{fig:strategy} clearly shows the transition between the $25 \times 3$ states of the DFTS.

The task is interpreted in the following natural language.
\begin{enumerate}
\item The swarm should infinitely visit ``goal" in ``triangle" formation, as long as ``battery'' is true.
\item All robots should avoid entering regions with obstacles.
\item The swarm should go back ``home'' to recharge once ``battery'' becomes false.
\end{enumerate}
It can be specified as an LTL formula $\psi \!=\! \psi_e \!\rightarrow\! \psi_s$, where $\psi_e := \square (\neg \mathrm{battery} \wedge  \mathrm{home} \rightarrow \bigcirc \mathrm{battery)} \wedge \square (\neg \mathrm{battery} \wedge  \neg \mathrm{home} \rightarrow \bigcirc \neg \mathrm{battery})$ and $\psi_s := \square \neg \mathrm{obstacle}
\wedge \square \lozenge (\mathrm{goal} \wedge \mathrm{triangle})
\wedge \square \lozenge \mathrm{battery}$.
The runtime safety requirements are formulated as bounded sets defined in Sec.~\ref{sec:qp} and encoded in the QP problem \eqref{eq: swarm_qp} for which the parameters can be found in our Github repository~\cite{ourCode}. We give two important parameters $T_{ud}\!=\!4\,$s and $u_{\max}\!=\!5\,$m/s.
The synthesis of the symbolic control strategy is solved using an off-the-shelf LTL toolbox, TuLiP~\cite{tulip2016}.
The QP problem is solved using the CasADi library~\cite{Andersson2019} with the ipopt solver on a commercial laptop with CPU i7-10750H.

The trajectories of the robots in a simulation run are shown in Fig.~\ref{fig:f1}. The swarm starts at ``home'' with the ``vertical'' formation. After leaving ``home'', the swarm goes along the horizontal corridor in the ``horizon'' formation, as shown in Fig.~\ref{fig:f11}, since the narrow space does not allow other formations. In Fig.~\ref{fig:f12}, the robot turns right and switches to the ``horizon'' formation to pass the short horizontal corridor. After reaching the ``goal'' in the open space, it switches to the ``triangle'' formation as specified by $\psi$, as shown in Fig.~\ref{fig:f13}. When any robots have low power (``battery'' signal is false) as shown in Fig.~\ref{fig:f14}, the swarm goes back to ``home'' to recharge. Once it gets charged at ``home'' and the ``battery'' signal is true again, the robot resumes its previous task to navigate itself to the ``goal'' again, as shown in Fig.~\ref{fig:f15}. From Fig.~\ref{fig:f1}, we can see that the controlled swarm ensures an obstacle-free trajectory when approaching the desired task. Also, proper formations are automatically switched to traverse narrow areas. The resulting behavior of the robot swarm completely satisfies the LTL specifications. This is also reflected by Fig.~\ref{fig:strategy} which visualizes the trajectory of the robot swarm in the abstract space. 
A video demonstration of this use case is accessible at {\small\url{https://www.youtube.com/watch?v=r1aecBOeDq0}}.

\begin{figure}[htbp]
\noindent
\hspace*{\fill}
\begin{tikzpicture}[scale=1,font=\scriptsize]
\node[anchor=south west] (eva) at (0cm, 0cm){\includegraphics[width=0.3\textwidth]{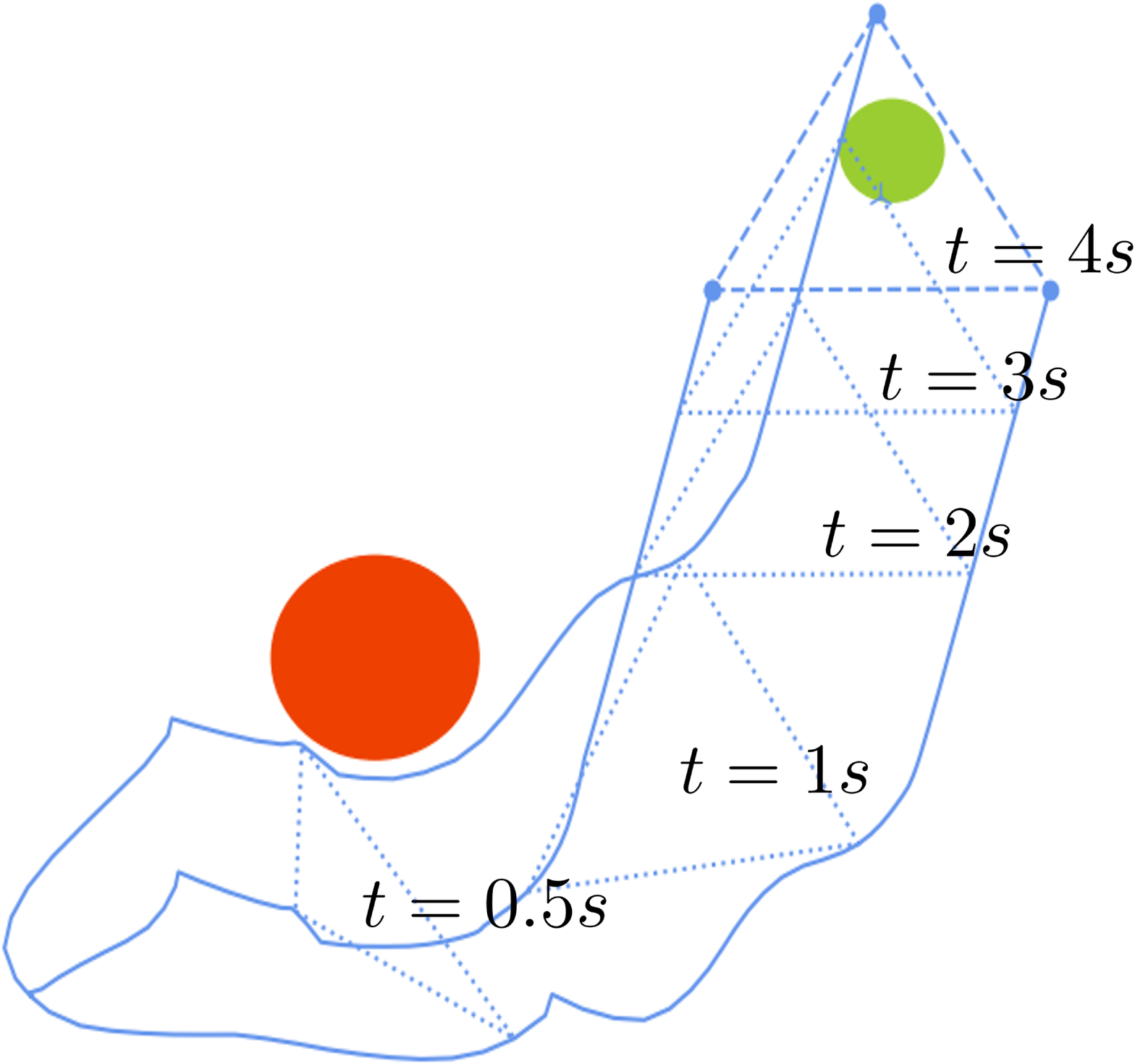}};
\node[circle,inner sep=0pt,minimum size=1.2mm,fill] (O) at (0.39cm,0.54cm) {};
\node[circle,inner sep=0pt,minimum size=1.2mm,fill,anchor=south east] (w) at (4.4cm,4.4cm) {};
\node[anchor=north east] () at (O) {$O$};
\node[] () at (2cm, 2.8cm) {Obstacle};
\node[anchor=south east] () at (w) {Waypoint};

\end{tikzpicture}
\hspace{\fill}
\caption{The satisfaction of the runtime safety requirements.}
\label{fig:formation}
\end{figure}

Fig.~\ref{fig:formation} indicates the satisfaction of runtime safety requirements, where the robot trajectories given a new waypoint and a desired formation are shown. In this case, a robot swarm from the initial position $O$ is required to achieve a ``triangle'' formation and its center simultaneously reaches the green rounded region within $T_{ud} \!=\! 4\,$s. During this period, all robots should avoid collision with the red rounded obstacle. In Fig.~\ref{fig:formation}, the trajectories of the robots are drawn as solid lines and the formation of the swarm is in dotted lines. It is shown that the robots successfully avoid the obstacle and finally reach the waypoint at a tolerable range. The ultimate formation is ``triangle'' and the reaching time is within $T_{ud}$. This study shows that the robot controller solved from the QP problem strictly ensures not only the runtime safety requirements but also the fixed-time convergence condition.

\section{CONCLUSION}\label{sec:con}
In this paper, we develop a formal-method-based framework for multi-robot swarm systems to design a reactive controller for the autonomous navigation task with automated formations. Under the synthesized symbolic controller for the abstract model, the QP-based control refinement approach can ensure that the behavior of the swarm system satisfies the LTL specification with runtime guarantees. In the future, the framework will be expected to extend to more robotic applications with more complicated specifications.



\balance

\bibliographystyle{ieeetr}
\bibliography{ref}

\end{document}

%% file: commands.tex
\newcommand{\X}{\mathbb{X}}

\newcommand{\A}{{A}} 
\newcommand{\ac}{a} 

\newcommand{\AP}{\mathsf{AP}}

\newcommand{\pia}{\mu}